\documentclass[final,5p,times,twocolumn]{elsarticle}
\usepackage{subfigure}
\usepackage{subscript}
\usepackage{graphicx,amsmath,amsfonts}
\usepackage{textcomp}

\usepackage{siunitx}
\usepackage{color}

\usepackage{lineno,hyperref}
\modulolinenumbers[5]
\journal{Journal of Magnetism and Magnetic Materials}

\usepackage[nogroupskip, nonumberlist, nopostdot, acronym]{glossaries}
\usepackage{color}
\newacronym{NO}{NO}{non-grain oriented}
\newacronym{GO}{GO}{grain oriented}
\newacronym{RD}{RD}{rolling direction}
\newacronym{TD}{TD}{transverse direction}
\newacronym{ND}{ND}{normal direction}
\newacronym{NA}{NA}{non-annealed}
\newacronym{SR}{SR}{stress-relieved}
\newacronym{SST}{SST}{Single Sheet Tester}
\newacronym{MSST}{Mini-SST}{Miniature Single Sheet Tester}
\newacronym{EF}{EF}{Epstein frame}
\newacronym{ODF}{ODF}{orientation distribution function}
\newacronym{EBSD}{EBSD}{electron backscatter diffraction}
\newacronym{RSST}{RSST}{Rotational Single Sheet Tester}
\newacronym{XRD}{XRD}{X-ray diffraction}
\newacronym{IEM}{IEM}{Institute of Electrical Machines}

\begin{document}

\begin{frontmatter}

\title{A New Approach to Measure Fundamental Microstructural Influences on the Magnetic Properties of Electrical Steel using a Miniaturized Single Sheet Tester}

\author[IEM]{N.~Leuning\corref{mycorrespondingauthor}} 
\ead{nora.leuning@iem.rwth-aachen.de}
\author[IMM]{M.~Heller}
\author[IEM]{M.~Jaeger}
\author[IMM]{S.~Korte-Kerzel}
\author[IEM]{K.~Hameyer}

\address[IEM]{Institute of Electrical Machines (IEM), RWTH Aachen University, D-52062 Aachen, Germany}

\address[IMM]{Institute for Physical Metallurgy and Materials Physics (IMM), RWTH Aachen University, D-52062 Aachen, Germany}

\cortext[mycorrespondingauthor]{Corresponding author}

\begin{abstract}
Magnetic properties of electrical steel are usually measured on Single Sheet Testers, Epstein frames or ring cores. Due to the geometric dimensions and measurement principles of these standardized setups, the fundamental microstructural influences on the magnetic behavior, e.g., deformation structures, crystal orientation or grain boundaries, are difficult to separate and quantify. In this paper, a miniaturized Single Sheet Tester is presented that allows the characterization of industrial steel sheets as well as from in size limited single, bi- and oligocrystals starting from samples with dimensions of \numproduct{10 x 22} \unit{\milli\meter}. Thereby, the measurement of global magnetic properties is coupled with microstructural analysis methods to allow the investigation of micro scale magnetic effects. An effect of grain orientation, grain boundaries and deformation structures has already been identified with the presented experimental setup. In addition, \textcolor{red}{a correction function is introduced to allow quantitative comparisons between differently sized Single Sheet Testers. This approach is not limited to the presented Single Sheet Tester geometry, but applicable for the comparison of results of differently sized Single Sheet Testers.} The results of the miniaturized Single Sheet Tester were validated on five industrial electrical steel grades. Furthermore, first results of differently oriented single crystals as well as measurements on grain-oriented electrical steel are shown to prove the additional value of the miniaturized Single Sheet Tester geometry. 
\end{abstract}

\begin{keyword}
Electrical steel \sep Miniaturized Single Sheet Tester \sep Single Crystals \sep Deformation \sep Grain boundaries \sep Electrical steel \sep FeSi
\end{keyword}

\end{frontmatter}

\nolinenumbers

\graphicspath{{figures/}}

\section{Introduction}

Magnetic properties of \gls{NO} and \gls{GO} electrical steels are usually obtained by measurements on standardized measurement sensors according to international standards, such as IEC-60404. For the macroscopic evaluation of the magnetic properties of electrical steel sheet this is sufficient. However, for a more detailed consideration of microstructural effects on the magnetic properties these standardized sensors are not suitable as the results are  insufficiently spatially resolved. In order to improve the understanding of the interrelations between grain orientation, grain boundaries and deformation mechanisms on the one hand and magnetic properties on the other hand, a miniaturized \gls{SST} was constructed and initial results are presented in this paper. 

Electrical steel sheet is used as magnetic core material for electrical machines, thus the magnetic properties are of main concern in the material selection \cite{Moses.2012}. Electromagnetic simulations are performed during the design stage of electrical machines to determine the relation between design, material choice and operational behavior of the machine. For such electromagnetic simulations, the magnetic properties of the electrical steel sheet material in question have to be modeled based on magnetic measurements of the magnetic permeability and iron loss \cite{Nell.2020}. These characteristic values are used to compare different electrical steel grades. With standardized measurement setups, the non-linear material behavior can be analyzed, different grades can be compared and iron loss models can be parameterized. Consequently, standardized measurements are crucial for the general application of electrical steel in electrical machines. For the development of improved electrical steel grades and advanced material models, advanced magnetic characterization approaches have to be utilized that go beyond standardized characterization techniques. Advanced methods that are used today include the consideration of vector characteristics of magnetic flux $B$ and magnetic field $H$, two-dimensional excitation conditions, rotating magnetic fields or local magnetic properties \cite{Yamagashira.2014,Lewis.2018,Thul.2018}. However, these techniques are mainly designed for polycrystalline sheet materials. In order to further study fundamental magnetization mechanisms of \gls{NO} and \gls{GO} steel another approach has to be developed that enables the quantification of effects on a grain scale. 

In order to study \textcolor{red}{individual} fundamental microstructure influences, we developed a \gls{MSST}. The minimum sample size is \numproduct{10 x 23} \unit{\milli\meter}. These sample dimensions allow the investigation of grown single crystals with specific orientations, grown bi-crystals with defined grain boundaries or oligocrystals with specifically adjusted deformation structures \cite{Heller.2021}. The measurements of the magnetic properties are not locally resolved within the \numproduct{10 x 17} \unit{\milli\meter} measurement area of the sample, but the Mini-\gls{SST} results are coupled with materials science microstructure investigation methods, i.e., hardness measurements, optical microscopy, X-ray diffraction, or \gls{EBSD} \cite{Heller.2021}. This approach allows a correlation  between crystallographic texture \cite{Yonamine.2004}, grain boundaries \cite{Campos.2006} and deformation mechanisms \cite{Landgraf.2000} with the magnetic properties. In addition, due to the small sample size, characterization of polycrystalline materials on a laboratory or industrial scale becomes possible even if the sample volume is small, e.g. during sample preparation of a manufactured motor.

\section{Miniaturization of a SST}

\subsection{Standardized Measurement Setups}
In general, three methods are used for the standardized
magnetic characterization of electrical steel, namely Epstein frames, \gls{SST} and ring core measurements. In this study, the \gls{SST} was miniaturized to allow the investigation of fundamental microstructural effects. In practice, none of the three standardized characterization methods outperformed the others, as they all have different advantages and disadvantages. Detailed information on these methods can be found in \textcolor{red}{DIN EN 60404 \cite{Norm}}, nevertheless a brief comparison is made here to illustrate the idea of the \gls{MSST}.

The measurement principle of all three methods can be summarized as follows: A magnetic field is generated by a current running through a copper winding (magnetization coil). The rectangular or ring shaped electrical steel sample is placed in this magnetic field and a secondary copper winding (induction coil) is placed as close as possible to the sample. The voltage, which is induced in this secondary induction winding is proportional to the flux density within the sample. Differences between the setups stem from the sample geometry on the one hand and the magnetic flux path on the other hand. In a ring shaped sample, the magnetic flux is closed entirely by the sample. In an Epstein frame,
four electrical steel strip legs are positioned in a rectangle with overlapping edges to close the magnetic flux path entirely by the sample material. In a \gls{SST}, there is only one sheet sample, which makes it impossible to close the magnetic path over the sample, thus, a double c-yoke is required. These macroscopically different magnetic flux paths lead to differences in the magnetic flux distribution across the sample cross sections. For example, due to the different magnetic path length at the outer and inner circumference of the ring cores and on the four legs of an Epstein frame, a flux concentration at the inner diameter occurs before the material is fully saturated. The resulting flux density distribution in the sample cross sections is inhomogeneous and thus, describes an averaged flux density within the sample. For the purpose of this study, it is necessary to have a homogenous flux density condition within the sample. In a \gls{SST}, the flux density distribution is homogeneous over the cross section. However, the yoke that is needed to close the magnetic flux can lead to additional losses and are determined by the yokes geometry. This is an effect that needs to be accounted for during the validation of the \gls{MSST} setup and is discussed in the following sections. Sample sizes for a \gls{SST} can be as large as \numproduct{500 x 500} \unit{\milli\meter}, whereas samples for the Epstein frame are \numproduct{280 x 30} \unit{\milli\meter}. The newly designed \gls{MSST} allows a minimum size of \numproduct{10 x 22} \unit{\milli\meter}.

\subsection{Mini SST - Geometry}

\begin{figure}[tbp] \centering
 	\centering
 	{\includegraphics[width=0.4\textwidth]{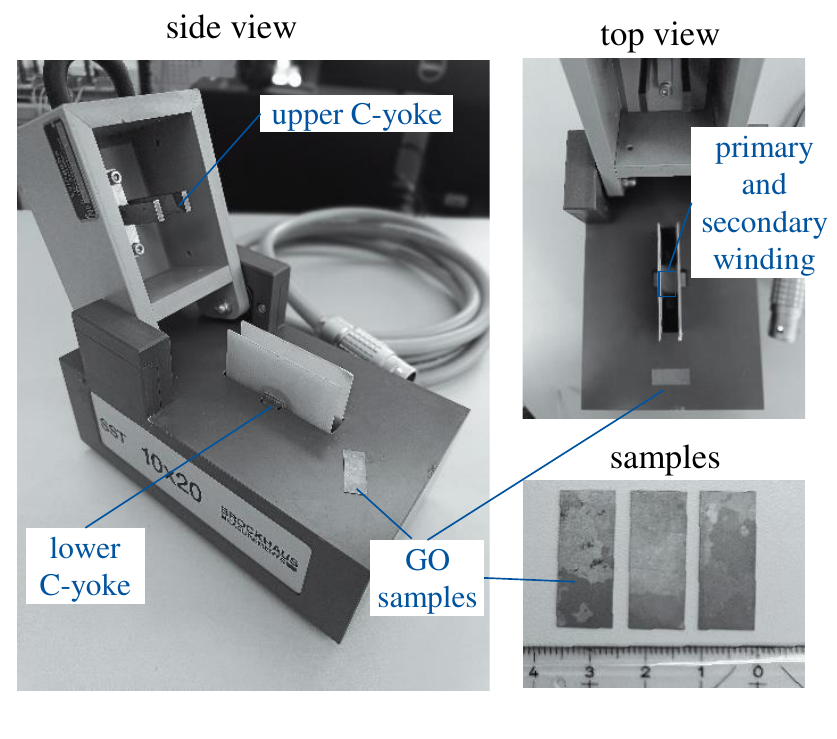}}
 	\caption{Photograph of the \numproduct{10 x 22} \unit{\milli\meter} Mini-SST.}
 	\label{fig:Mini-SST}
 \end{figure}

The purpose of the \gls{MSST} is to enable a detailed characterization of the magnetic properties of single, bi- and oligo-crystals with dimensions achievable by crystal growth \cite{Heller.2021}. Since the size of samples produced by crystal growth are in the range of \SI{2}{\square\centi\meter}, a correspondingly small \gls{SST} was developed in cooperation with \textit{Brockhaus Measurements}. The outer distance between the magnet yoke legs of the Mini-SST is about \SI{22}{\milli\meter} defining the minimum sample length, however, the free magnetic path length $l_{\text{m}}$ between the yoke poles is about \SI{16}{\milli\meter}. Consequently, the pole thickness is \SI{3}{\milli\meter} on each side. Both, the primary winding $N_{1}$ and secondary winding $N_2$ have 60 turns each. The \gls{MSST} is controlled by an \textit{MPG 200} test bench from \textit{Brockhaus Measurements}. A picture of the \gls{MSST} is displayed in Fig.~\ref{fig:Mini-SST}.

\textcolor{red}{Magnetic properties are measured by means of electric measurements, i.e., electric current and voltage. The required current is supplied by a power amplifier. Thereby, a magnetic field is created by the primary magnetization winding. The current is measured by means of a temperature-stable, low inductivity precision resistor. Polarization is determined by measuring the induced voltage in the secondary induction winding. The parallel recording of magnetic field $H$ and magnetic polarization $J$ with separate analogue-digital converters enables simultaneous measurement. A control algorithm is used to ensure sinusoidal excitation, where the secondary voltage can be checked and constantly regulated in accordance with the nominal value. The nominal voltage is supplied by a highly stable, digital frequency generator. Amplitude and frequency are set by software according to the sample data entered and the default values, according to the \textit{MPG 200} and \textit{MPG Expert} Software.}

\section{Description of the Correction Function}
It is well known that the characterization of magnetic properties strongly depends on the geometric conditions of the measurement setup, e.g., Epstein frame to \gls{SST} as well as differently sized \gls{SST}, which are summarized in \cite{Sievert.2000}. Even different laboratories with similar geometric conditions, as presented in \cite{Sivert.2011} for Epstein measurements, come to different results. Therefore, the comparability between measurement results becomes a matter of reference samples as well as measurement and correction techniques. To ensure comparability between the \gls{MSST} measurements and previous measurements of the authors, a general correction function is developed and parameterized to the reference \gls{SST} of the \gls{IEM}. 

\begin{figure}[tbh] \centering
	\centering
	\subfigure[Low magnetic fields.]
	{\includegraphics[]{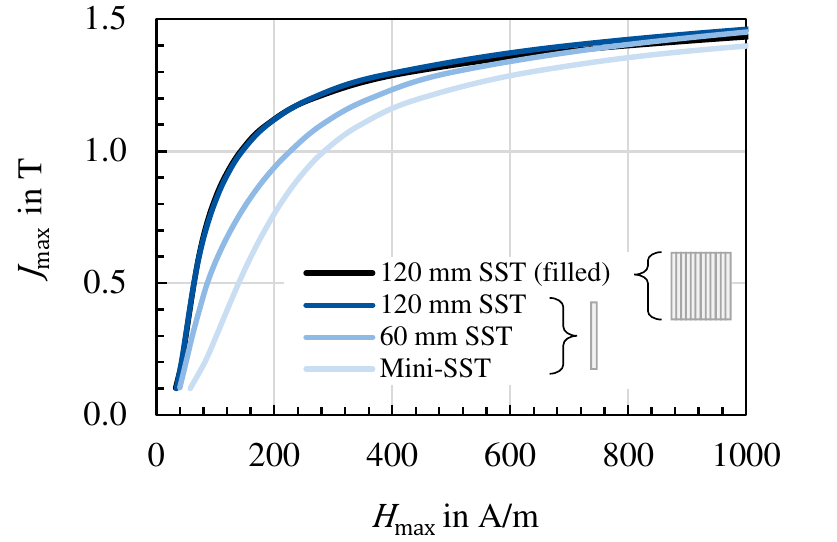}}
	\subfigure[High magnetic fields.]
	{\includegraphics[width=0.44\textwidth]{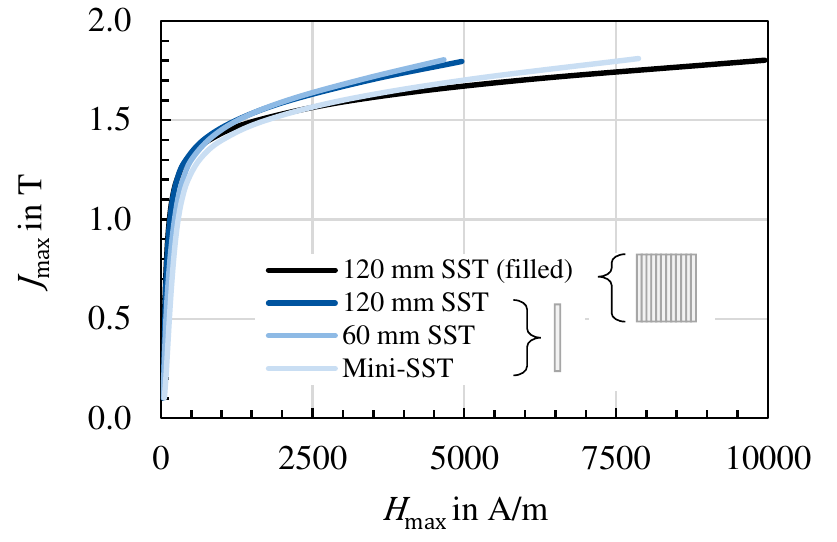}}	
	\caption{Measurements of a M330-50A (M1) reference sample of $d_{\text{strip}} =$  \SI{1}{\centi\meter} on differently sized \gls{SST}s for the low (a) and high magnetic field region (b) and a measurement of 12 strips resp. of $d_{\text{strip}} =$  \SI{1}{\centi\meter}. }
	\label{fig:Magnetization_curves}
\end{figure}

In Fig.~\ref{fig:Magnetization_curves}, magnetization curves tested at \SI{50}{\hertz} are displayed for three differently sized \gls{SST}s of \numproduct{120 x 120} \unit{\milli\meter}, \numproduct{60 x 60} \unit{\milli\meter} and \numproduct{10 x 20} \unit{\milli\meter}. To ensure comparability, the same sample was measured on all \gls{SST}s. A strip of $d_{\text{strip}} =$ \SI{10}{\milli\meter} and $l_{\text{strip}} =$ \SI{120}{\milli\meter} was utilized for this. To account for the cut-edge effect, the sample for the filled \numproduct{120 x 120} \unit{\milli\meter} \gls{SST} consisted of 12 sample strips each with a width of \SI{10}{\milli\meter}  and a length of \SI{120}{\milli\meter} that are taped together according to \cite{Schoppa.2000}. When looking at the magnetization curves, two effects can be observed. At low magnetic fields $H$, as can be seen in Fig.~\ref{fig:Magnetization_curves} (a), the smaller \gls{SST}s generally show lower magnetization compared to larger \gls{SST}s. Furthermore, the curves for one \SI{10}{\milli\meter} sample strip are identical to those for a fully filled \gls{SST} (12 $\times$ \SI{10}{\milli\meter}) at low magnetic fields. At high magnetic fields $H$, the fully filled reference \gls{SST} shows the hardest magnetization behavior right before the \numproduct{10 x 20} \unit{\milli\meter} \gls{SST}. Both, the  \numproduct{60 x 60} \unit{\milli\meter} \gls{SST} and the reference \gls{SST} with just one sample strip need much lower field strengths to seemingly magnetize the sample to \SI{1.8}{\tesla}.  

The observed behavior is attributed to two separate effects. The first one is linked to the magnetic resistance of the yoke, whereas the second effect is linked to the influence of stray flux within the unfilled coil cross section.
Due to the required size of the yoke's pole width, non-ideal conditions in the ratio of the free magnetic path length and the yoke height occur, when the \gls{SST} is downscaled. Thereby, the permeability of the yoke becomes an important factor. Moreover, at high magnetic fields, the air flux needs to be considered. The free space in the coil depends on the solenoid housing, which is fixed and the respective sample cross section (i.e. sample thickness and width). This explains the strong difference between the results for the filled and non-filled reference \gls{SST}, as well as for the differently sized \gls{SST}s with different solenoid housing.

To account for these differences and ensure comparability, a correction function for the $B(H)$ measurement results has been developed and parametrized to the \numproduct{120 x 120} \unit{\milli\meter} \gls{IEM} \gls{SST}, which, as previously stated, serves as the reference \gls{SST} within this study. Again, the correction function for the \gls{MSST} is necessary to account for the magnetic resistance of the yoke and the air flux in the solenoid, which enables a quantitative comparison to the reference \gls{SST}. The proposed method describes a general approach, that can be transferred to other research facilities and \gls{SST} sizes to improve comparability between measurement setups and research facilities. 

\begin{figure}[tbp] \centering
	\centering
	{\includegraphics[]{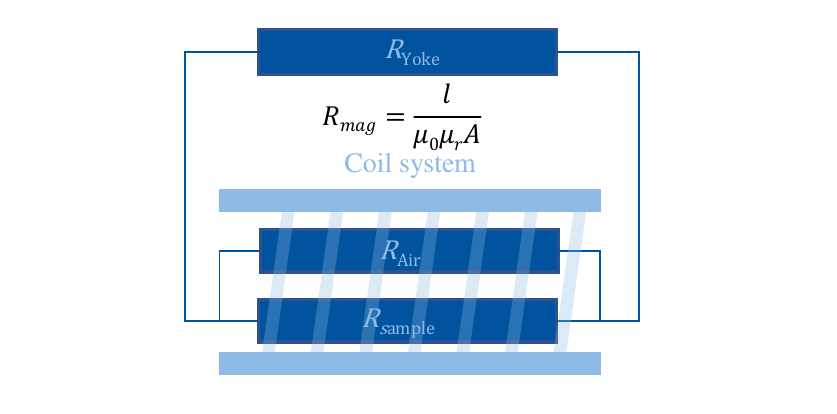}}
	\caption{Magnetic equivalent circuit framework to calculate a \gls{SST} correction function.}
	\label{fig:Equivalentcircuit}
\end{figure}

For a designated reference \gls{SST}, certain assumptions have to be made. Firstly that the reference \gls{SST} is ideal and only measures the resistance of the sample. Hence, there is no stray flux outside the yoke, which has a permeability $\mu_{\text{r}} = \infty$. Therefore, the magnetic resistance is only composed of the resistance of the sample and the air in the coil. As a framework for the correction function a magnetic equivalent circuit is used (Fig.~\ref{fig:Equivalentcircuit}), where:

\begin{itemize}
	\item Mini-SST geometry is known: $A_{\text{yoke}}$ $l_{\text{yoke}}$
	\item Sample geometry is known: $A_{\text{sample}}$ $l_{\text{sample}}$
	\item Magnetic flux is known: $\Phi =I \cdot N$ and 
	\item Total flux linkage $\Psi = \Psi_{\text{Air}} + \Psi_{\text{Sample}}$
\end{itemize}

with $A$ representing a cross section, $l$ representing a length, $I$ describing an electric current and $N$ the number of turns.
 The magnetic resistance can generally be calculated using the following equation:
\begin{equation}
	R_{\text{mag}}=\frac{l}{\mu_0 \mu_{\text{r}} A} ~.
\end{equation}

\subsection{Correction of the Magnetic Field Strength}

The measured magnetization curves $B_{\text{meas.}}(H_{\text{meas.}})$ of the Mini-SST cannot be directly compared to the reference, as previously stated, thus a corrected $B_{\text{meas., corr.}}(H_{\text{meas., corr.}})$ has to be calculated. Inputs for the calculation of the corrected magnetic field 	$H_{\text{meas.,corr.}}$ are the measured flux density $B_{\text{meas.}}$ in T and the measured magnetic field strength $H_{\text{meas.}}$ in A/m.  With the following equations the corrected field strength can be determined.

The total flux linkage $\Psi$ can be calculated from the flux density $B_{\text{meas.}}$ in the sample as given by the measurement system  and the cross section of the sample $A_{\text{sample}}$, which is determined geometrically (thickness $x$ width).
\begin{equation}
	\Psi= B_{\text{meas.}} \cdot A_{\text{sample}}
	\label{eq:psi}
\end{equation}

The magnetic flux $\Phi$ is the product of the measured magnetic field strength $H_{\text{meas.}}$ and the magnetic path length $l_{\text{m}}$ between the yoke legs of the \gls{MSST} according to
\begin{equation}
	\Phi= H_{\text{meas.}} \cdot l_{\text{m}} ~.
	\label{eq:phi}
\end{equation}

With the information of $\Psi$ (eq. (\ref{eq:psi})) and the cross section of the yoke pole surfaces $A_{\text{yoke}}$, the magnetic flux density in the yoke $B_{\text{yoke}}$ can be calculated
\begin{equation}
   B_{\text{yoke}}=\frac{\Psi}{A_{\text{yoke}}}
\end{equation}

\begin{figure}[tbp] \centering
	\centering
	{\includegraphics[]{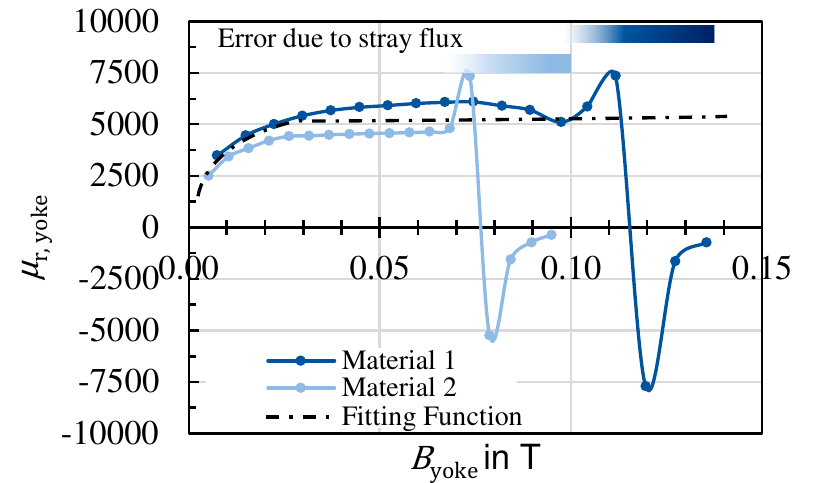}}
	\caption{Determination of the fitting function to account for the permeability of the \gls{MSST} yoke.}
	\label{fig:permeability}
\end{figure}

For the calculation of  the magnetic field in the yoke $H_{\text{yoke}}$, the permeability of the yoke needs to be determined. As this value cannot be directly measured, a fitting has been performed. As depicted in Fig.~\ref{fig:permeability}, the permeability of the yoke is fitted to  $B_{\text{yoke}}$ measurement values below \SI{0.07}{\tesla} of two materials and can be described by the following empirical equation:

\begin{equation}
	\mu_{\text{r, yoke}} = 	\frac{7750}{1 + 2^{-160 \cdot B_{\text{yoke}}}}-2000
	\label{eq:mu}
\end{equation}

\begin{equation}
	H_{\text{yoke}}=\frac{B_{\text{yoke}}}{\mu_{\text{r, yoke}} \cdot \mu_{\text{0}}}
	\label{eq:Hyoke}
\end{equation}

With equations (\ref{eq:psi}) to (\ref{eq:Hyoke}), all the variables needed to calculate the corrected magnetic field taking the yoke permeability into account are known, resulting in the following equation: 

\begin{equation}
	H_{\text{meas.,corr.}}=\frac{1}{l_{\text{m}}} \cdot (\Phi - H_{\text{yoke}} l_{\text{yoke}}) ~.
\end{equation}

\subsection{Correction of the Magnetic Flux Density}
 
 To account for the influence of stray flux, $	H_{\text{stray}}$ (\ref{eq:Hstray}) and $	B_{\text{stray}}$ (\ref{Bstray}) are calculated with the help of the magnetic flux $\Phi$ and the properties of the yoke.
  These parameters are used to subsequently determine the flux $\Psi_{\text{stray}}$ and $\Psi_{\text{sample}}$ in equations (\ref{eq:psistray}) and (\ref{eq:psisample}). 
 
 \begin{equation}
 	H_{\text{stray}}=\frac{1}l_{\text{stray}} \cdot (\Phi-H_{\text{yoke}} l_{\text{yoke}})
 	\label{eq:Hstray}
 \end{equation}

\begin{equation}
	B_{\text{stray}} = 	H_{\text{stray}} \cdot \mu_{0}
	\label{Bstray}
\end{equation}

\begin{equation}
	\Psi_{\text{stray}}= 	B_{\text{stray}} \cdot A_{\text{coil}}
	\label{eq:psistray}
\end{equation}
 
\begin{equation}
	\Psi_{\text{sample}}= 	\Psi - \Psi_{\text{stray}}
	\label{eq:psisample}
\end{equation}

The value for the length of the magnetic stray lines $l_{\text{stray}}$ is based on the assumption that the flux lines are longer than the direct connection between the poles of the magnetic yoke $l_{\text{m}}$, which is \SI{16}{\milli\meter}. A value of $+12.5\%$ is assumed, which, in our case, corresponds to \SI{18}{\milli\meter}. The free cross section of the solenoid $A_{\text{coil}}$ is approximated with the help of geometrical measurements to be \SI{42}{\square\milli\meter}. Finally, the corrected flux density $B_{\text{meas. corr}}$ in the sample can be determined according to (12):

\begin{equation}
	B_{\text{meas. corr.}}= \frac{\Psi_{\text{sample}}}{	A_{\text{sample}}}~.
\end{equation}

 \begin{figure}[bht] \centering
	\centering
	\subfigure[Low magnetic fields.]
	{\includegraphics[]{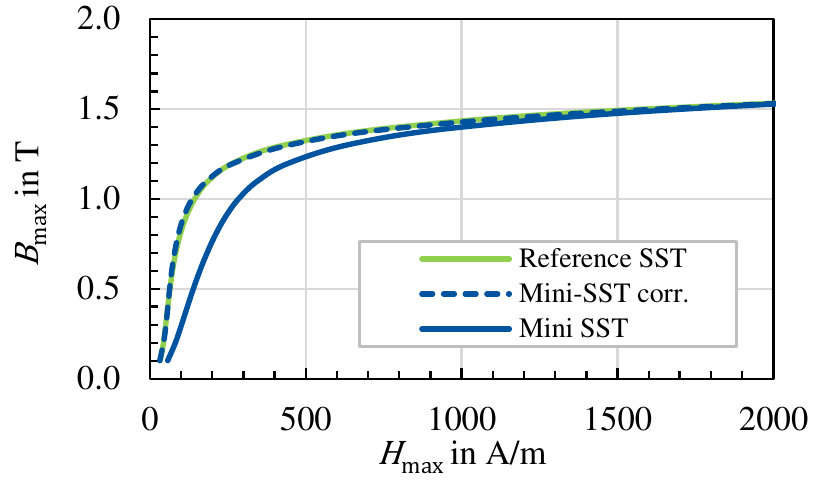}}
	\subfigure[High magnetic fields.]
	{\includegraphics[]{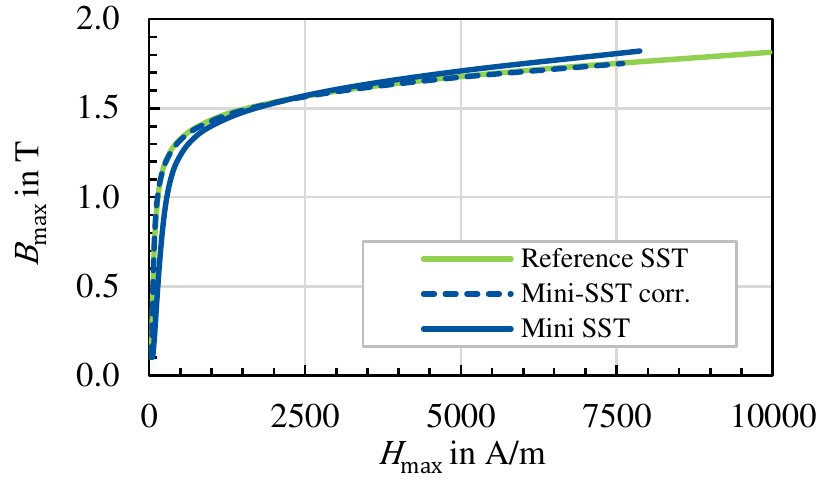}}	
	\caption{Measurements and corrected $B(H)$-curves of Material 1 (M330-50A).}
	\label{fig:Magnetization_curves_corrected}
\end{figure}

In Fig.~\ref{fig:Magnetization_curves_corrected} results of the corrected $B(H)$-curves are displayed in comparison to the uncorrected measurements on the \gls{MSST} and measurements of the same material on the reference \gls{SST}. It is evident that the identified and fitted parameters of the correction function lead to a measurement data correction that enables a quantitative comparison between the differently sized \gls{SST}s.  In both the low and high $H$ region, the curves for the reference and corrected  $B(H)$ characteristics are virtually congruent.

\section{Validation of the Correction Function}

\begin{figure}[bp] \centering
	\centering
	{\includegraphics[]{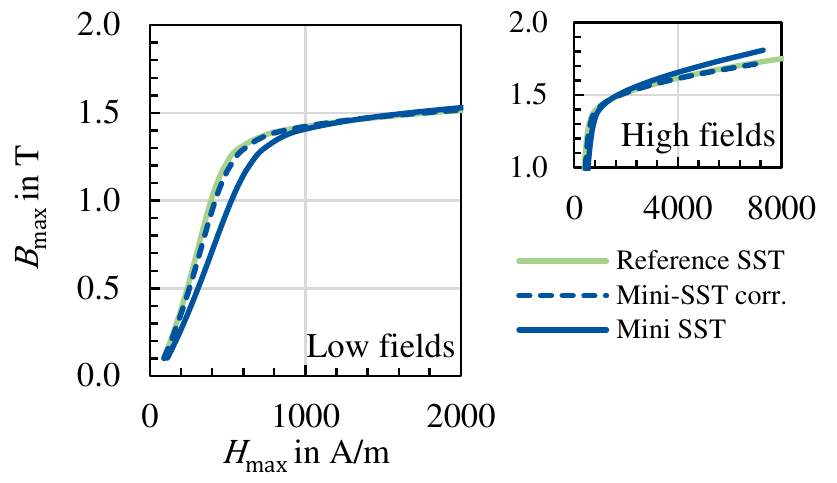}}
	\caption{Measurements and corrected $B(H)$-curves of Material 2.}
	\label{fig:Material_2}
\end{figure} 

\begin{figure}[bp] \centering
	\centering
	{\includegraphics[]{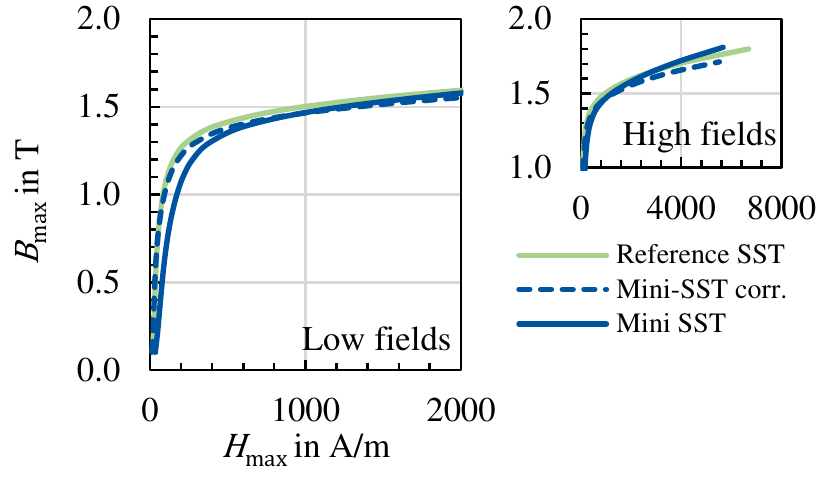}}
	\caption{Measurements and corrected $B(H)$-curves of Material 3.}
	\label{fig:Material_3}
\end{figure} 

\begin{figure}[bp] \centering
	\centering
	{\includegraphics[]{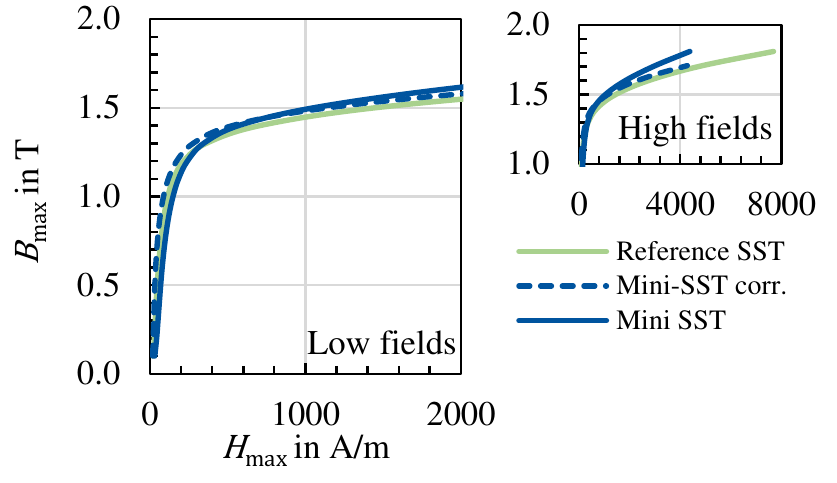}}
	\caption{Measurements and corrected $B(H)$-curves of Material 4.}
	\label{fig:Material_4}
\end{figure} 

\begin{figure}[bp] \centering
	\centering
	{\includegraphics[]{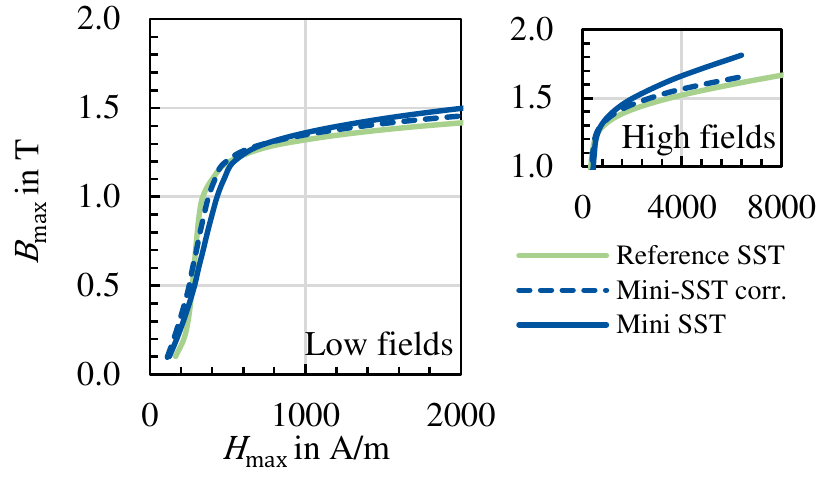}}
	\caption{Measurements and corrected $B(H)$-curves of Material 5.}
	\label{fig:Material_5}
\end{figure}

In order to validate the correction function, the approach is tested on a total of five different materials with different chemical compositions, grain sizes, sheet thicknesses and resulting magnetic properties. An overview of the tested materials is given in Table~\ref{tab:Material_1}.

\begin{table}[h]
	\centering
	\begin{tabular}{lcccc}     
		~  & $d_{\text{sheet}}$ & orientation &  $d_{\text{GS}}$   \\
		\hline    
		Material M1 & \SI{0.50}{\milli\meter} & NO, RD & 52 $\muup$m  \\ 
		Material M2 & \SI{0.35}{\milli\meter} & NO, RD  & 69 $\muup$m \\
		Material M3 & \SI{0.27}{\milli\meter} & NO, RD & 134 $\muup$m \\ 
		Material M4 & \SI{0.20}{\milli\meter} & NO, RD  & 95 $\muup$m \\
		Material M5 & \SI{0.18}{\milli\meter} & GO, TD  & \SI{1}{\centi\meter} \\ 
		\hline
	\end{tabular}
	\caption{Nominal thickness $d_{\text{sheet}}$, orientation and mean grain diameter $d_{\text{GS}}$ of the studied materials.}
	\label{tab:Material_1}
\end{table}

In Fig~\ref{fig:Material_2} to Fig.~\ref{fig:Material_5} the results are displayed for all materials. In general, the correction function improves the compatibility for all materials, however, the results for M1 show the best result. \textcolor{red}{This could be due to the small grain size, together with the effect of crystal orientation and resulting stray fields. This is a topic that can further be studied with the validated \gls{MSST}}. The correction function can also be applied to \gls{GO} material, which helps the interpretation of results for single crystals due to the large grains and variable sample orientation. As the purpose of the \gls{MSST} is the characterization of samples that cannot be tested on a reference \gls{SST} due to their size, the function cannot be parametrized on each material but must be generally applicable. The results presented suggest that a quantitative comparison is enabled with sufficient accuracy.

\section{\textcolor{red}{Measurements of undeformed single crystals and GO material}}

In this section, preliminary results of \gls{MSST} measurements of single crystal as well as \gls{GO} material coupled with \gls{EBSD} measurements are presented. 

\begin{figure}[ptb] \centering
	\centering
	\subfigure[Low magnetic fields (IEM \gls{MSST} results).]
	{\includegraphics[width=0.44\textwidth]{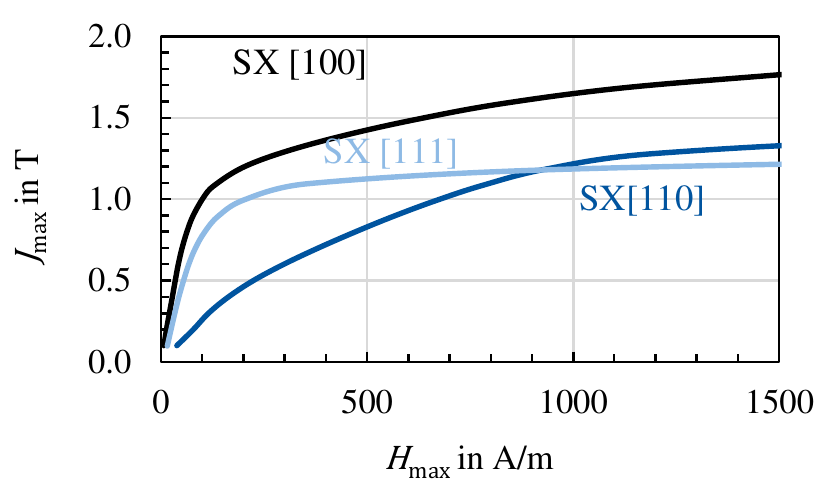}}
	\subfigure[High magnetic fields comparison of IEM and Honada et al. results.]
	{\includegraphics[width=0.44\textwidth]{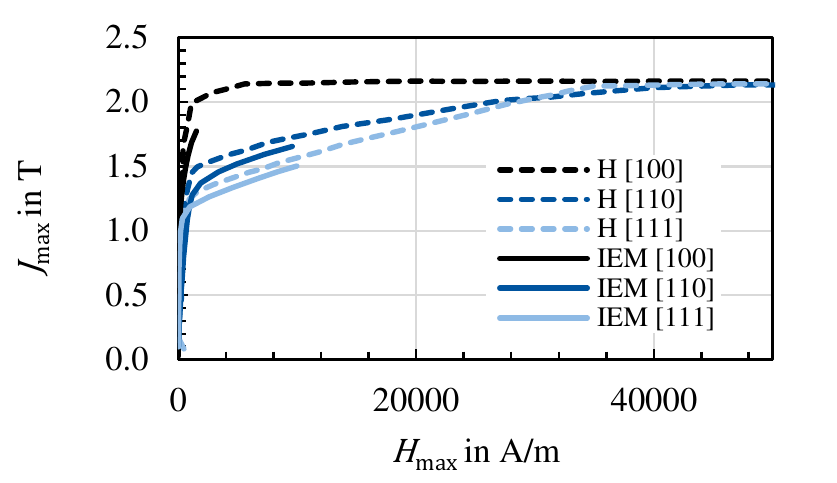}}
	\subfigure[Low magnetic fields (after \cite{HONDA.1926}.]
	{\includegraphics[width=0.44\textwidth]{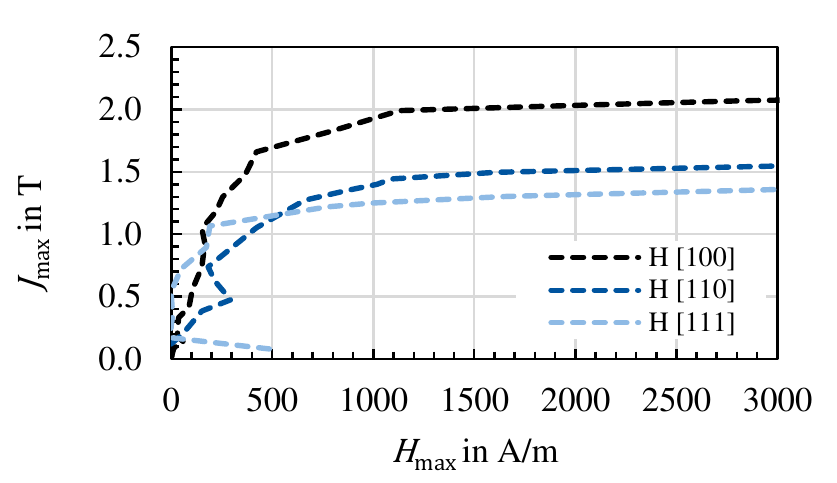}}
	\caption{Results of single crystal measurements and comparison to literature data \cite{HONDA.1926}.}
	\label{fig:Single_Crystals}
\end{figure}

\textcolor{red}{The single crystals were produced on a self build induction furnace that works according to the Bridgman Stockbarger method. Subsequently, the sheet geometry, necessary for the \gls{MSST}, was cut out of the cylindric crystal growth geometry with the help of an electrical discharging machine. To make sure that no heat affected layer or deformation layer remains the sample was etched with nitric acid (\SI{100}{\milli\liter} HNO3, \SI{150}{\milli\litre} H2O), mechanically grinded and polished and the final surface finish for EBSD was achieved with electro polishing (A2 without water for \SI{15}{\second} at \SI{24}{\volt}). A detailed description can be found in \cite{Heller.2021,Heller.08.02.2023}.} The \gls{GO} material was cut from industrial transformer sheets with a strong Goss texture along different orientations relative to \gls{RD}.

The single crystals were characterized on the \gls{MSST} at \SI{50}{\hertz} with peak inductions between \SI{0.1}{\tesla} and \SI{1.5}{\tesla} in \SI{0.1}{\tesla}-steps. The magnetization curves $J(H)$ are shown in Fig.~\ref{fig:Single_Crystals} (a). The magnetic anisotropy of the three common axes is clearly visible. Magnetization of the [100] single crystal is easiest, as expected. At first sight, the curves for the [110] and [111] single crystals show an unexpected behavior as the magnetization in [110] seems to be harder compared to the [111] direction. In order to examine this behavior further, the results have been compared to the data of Honda et al. \cite{HONDA.1926} and their work on the magnetization behavior of single crystalline iron, which is shown in Fig.~\ref{fig:Single_Crystals} (b). It can be seen that the results actually show similar behavior when looking at higher field strengths. Due to the different measurement setup, Honda was able to measure higher field strength up to saturation. The slight differences of the curves can stem from the difference in chemical composition as well as the manufacturing and preparation of samples. Looking at the low magnetic field region, as displayed in Fig.~\ref{fig:Single_Crystals} (c) the same initial crossing of the [110] and [111] curves can be observed, which could have been due to measurement scatter of Honda's results but shows a smooth transition for the \gls{MSST} measurements. In ongoing work, the single crystals, from which the here analyzed samples have been cut, are subsequently plastically deformed in several steps. After each step, one sheet sample is cut for \gls{MSST} measurements. This is an additional advantage of the \gls{MSST} setup as all deformed samples can be cut out of the same single crystal \textcolor{red}{ \cite{Heller.08.02.2023}}. 

   \begin{figure}[tb] \centering
   	\centering
   	\subfigure[\textcolor{red}{EBSD measurements of the pole figures and exemplary sample preparation.}]
   	{\includegraphics[]{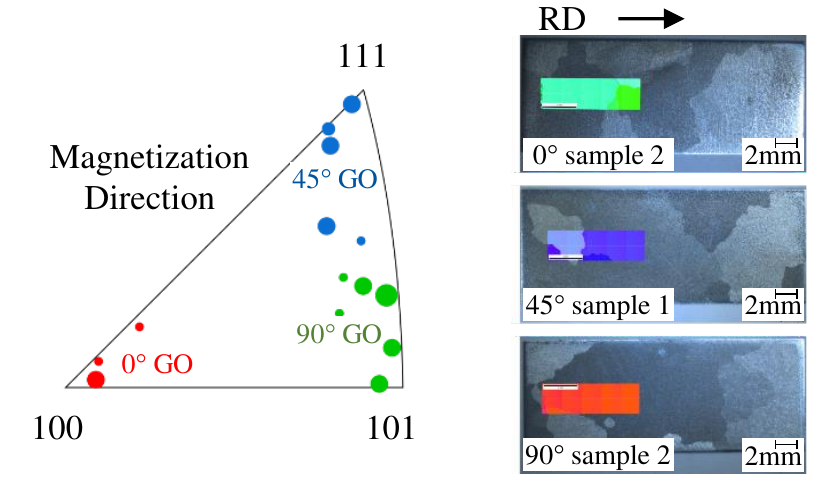}}
   	\subfigure[Magnetization curves at  \SI{50}{\hertz}.]
   	{\includegraphics[]{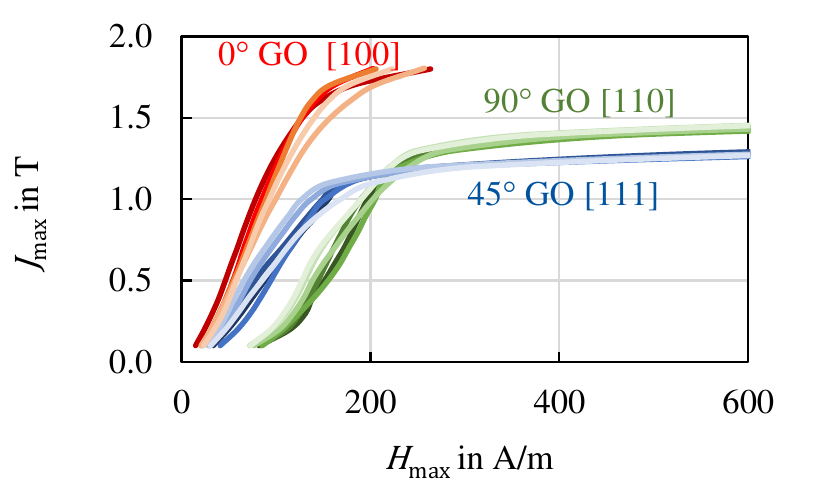}}
   	\caption{Results of industrial \gls{GO} sheet measurements.}
   	\label{fig:GO}
   \end{figure}

In a second example, results of industrial \gls{GO} \gls{MSST}  measurements are displayed in Fig.~\ref{fig:GO}. Six samples per sheet direction in \gls{RD} (\ang{0}), \gls{TD} (\ang{90}) and diagonal sheet plane direction (\ang{45}) have been cut from industrial \Gls{GO} sheet. Due to the crystallographic orientation of the present Goss grains, the sample directions along magnetization direction correspond  to the [100], [110] and [111] direction of the unit cells. \gls{EBSD} measurements were performed on the samples to validate the crystallographic orientation. Inverse pole figures, as depicted in Fig.~\ref{fig:GO} (a), show which crystallographic plane normal is parallel to a particular sample axis, in this case the rolling direction. Therefore, points in the triangle show the exact orientation of these crystallographic plane normals relative to the rolling direction and the colour groups them into near 100 (red), 110 (green) and 111 (blue). The corresponding magnetization curves are shown in Fig.~\ref{fig:GO} (b). These results correspond very well to the single crystal measurements in Fig.~\ref{fig:Single_Crystals} (a).
\textcolor{red} {Both, \gls{GO} and the single crystals have pronounced crystal orientations. With a \gls{SST}, magnetic field $H$ and magnetic polarization $J$ are treated as scalar properties, although they are actually vectors. For the [100] single crystal and \gls{GO} cut in {RD}, the vector and scalar values are expected to be equal, as the easy magnetization direction, an thus, orientation of magnetic domains are aligned parallel with the magnetic field generated perpendicular to the magnetizing coil. For differently oriented single crystals and \gls{GO} which is cut in unfavourable directions, this is not the case and the domains want to align in the easy directions, which are not parallel to the induced magnetic field. Only the polarizations vector component parallel to the magnetic field is obtained. When the sample approaches saturation, the domains are forced out of the easy directions into the direction of the applied magnetic field. As a result, the mismatch of vector and scalar properties decreases at high polarizations. The systematic error of neglecting vector properties needs to be accounted for in the evaluation and interpretation of results of \gls{SST} measurements in general. However, this systematic error is inherent of the \gls{SST} setups and its scalar consideration of vector properties. As the approach wants to enable a comparison between different \gls{SST}, the fundamental measurement principle is not changed.}

 In the examples given, the chemical composition and sample preparation are different, nevertheless both examples show how the influence of orientation can be analyzed without the influence of high angle grain boundaries. In ongoing work, bi-crystals have been successfully grown and subsequently deformed to study the effect of high angle grain boundaries as well as their deformation behavior. In the future, these measurements may be correlated in more detail with microstructural parameters that can be controlled and quantified in such small samples, such as dislocation density and domain distribution. The measurements and validations shown here highlight that small scale characterisation is a promising method to better understand the characteristics of microstructural parameters  of electrical steel.

\section{Conclusions}
In this paper a miniaturized \gls{SST} is presented that has been designed to study fundamental microstructural effects on the magnetic properties of electrical steels. A correction function is developed to account for the non-ideal geometric conditions of the setup, i.e., the air flux in the solenoid and the yoke pole to sample ratio, to allow a comparison to a reference \Gls{SST}. A validation of the correction function is performed on five industrial materials. The  results of this paper can be summarized in the following points:             

\begin{itemize}
	\item The validation shows that measurement results of the \gls{MSST} can be quantitatively compared to those established reference \gls{SST} setups, after a one-time parametrization on industrial steel sheet.
	
	\item The correction function approach can be transferred to other \gls{SST} setups, as it mainly depends on geometric conditions and a measurement at low magnetic fields to determine a fitting function for the permeability of the yoke. 
	
	\item Fundamental micro magnetic effects of orientation, deformation and grain boundaries can be assessed as sample the minimum sample size allows the analysis of grown single-, bi- and oligo crystals. Additional microstructural analysis is necessary to link the effects to the magnetic results.
	
	\item In case of industrial \gls{NO} and \gls{GO} material, the \gls{MSST} can be useful in cases where sample material is sparse, i.e., experimentally produced laboratory grades or materials of manufactured machines.
\end{itemize}
	
\textcolor{red}{With the \gls{MSST} and parametrized correction function, studies of the microstructure influences and of industrial \gls{NO} can be performed analogous to usually sized \gls{SST}. A quantitative comparison is enabled. The challenge in the transfer of the correction function to various sized \gls{SST} lies in the determination of the geometric parameters and especially in the assumption of the stray field length. The value cannot be directly measured so needs to be fitted empirically or possibly simulated.}

\section*{Acknowledgement}
The \gls{MSST} was funded by the Deutsche Forschungsgemeinschaft
(DFG, German Research Foundation) as part of the DFG-research group –
"FOR 1897 – Low-Loss Electrical Steel for Energy-Efficient Electrical Drives".

\section{References}

\bibliography{mybibfiles}

\end{document}